\documentclass{amsart}
\usepackage[dvips]{color}
\usepackage{amssymb,amsmath,amscd, amsthm,stmaryrd,verbatim, mathrsfs}

\input prepictex
\input pictex
\input postpictex

\newtheorem{Th}{Theorem}
\newtheorem{theorem}{Theorem}[section]

\newtheorem{Thm}[theorem]{Theorem}

\newtheorem{Def}[theorem]{Definition}

\theoremstyle{definition}
\newtheorem{Ques}[theorem]{Problem}

\theoremstyle{remark}
\newtheorem{Rmk}[theorem]{Remark}
\newtheorem{Fact}[theorem]{Facts}

\numberwithin{equation}{section}

\newcommand{\mb}{\mathbf}
\newcommand{\bb}{\mathbb}
\newcommand{\ms}{\mathscr}
\newcommand{\mr}{\mathrm}
\newcommand{\frk}{\mathfrak}

\usepackage{hyperref}
\begin{document}

\title{The Representation Aspect of the Generalized Hydrogen Atoms }

\author{Guowu Meng}

\address{Department of Mathematics, Hong Kong Univ. of Sci. and
Tech., Clear Water Bay, Kowloon, Hong Kong}

\email{mameng@ust.hk}

\subjclass[2000]{Primary 22E46, 22E70; Secondary 81S99, 51P05}

\date{April 23, 2007}


\keywords{conformal groups, generalized hydrogen atoms, Wallach
representations.}

\begin{abstract}
Let $D\ge 1$ be an integer. In the Enright-Howe-Wallach
classification list of the unitary highest weight modules of
$\widetilde{\mr{Spin}}(2, D+1)$, the (nontrivial) Wallach
representations in Case II, Case III, and the mirror of Case III are
special in the sense that they are precisely the ones that can be
realized by the Hilbert space of bound states for a generalized
hydrogen atom in dimension $D$. It has been shown recently that each
of these special Wallach representations can be realized as the
space of $L^2$-sections of a canonical hermitian bundle over the
punctured ${\bb R}^D$. Here a simple algebraic characterization of
these special Wallach representations is found.
\end{abstract}
\maketitle \tableofcontents

\section{Introduction}
The generalized hydrogen atoms, discovered in the late 60s by
McIntosh and Cisneros \cite{MC70} and independently by Zwanziger
\cite{Z68}, are hypothetic atoms where the nucleus carries both
electric and magnetic charges. Their extension to dimension five
were obtained by Iwai \cite{Iwai90} in the early 90s, their
construction and preliminary analysis in all dimensions higher than
or equal to three were given about two years ago by this author
\cite{meng05}, and their extension to dimensions one and two will be
given in appendix \ref{Appendix1} of this paper.

The main purpose here is to elaborate on the representation
theoretical aspect of the generalized hydrogen atoms on the one hand
and to give a simple algebraic characterization of a special family
of Wallach representations on the other hand. The message I wish to
convey to mathematical physicists is that the generalized hydrogen
atoms are mathematically beautiful, and the message I wish to convey
to mathematicians is that, for the (spin-)conformal group of the
(compactified) Minkowski spaces, the Wallach representations in Case
II, Case III, and the mirror of Case III  from the classification
list of Ref. \cite{EHW82} admit a very simple algebraic
characterization.

For readers who are only interested in mathematics, theorems 1 and 2
below are our main mathematical results, theorem 3 below can be
skipped, and any paragraph involving the phrases such as
``generalized hydrogen atoms" or ``MICZ-Kepler problems" can be
ignored; for example, the entire appendix A can be ignored. In other
word, this is a mathematical paper which is rigorous by the current
mathematical standard, but it is motivated by physical problems and
it enhances our understanding of the physical models.

To state the main results, we need to first recall some basic facts
and introduce some notations.

\subsection{Pseudo-orthogonal groups} Let $p$, $q$ be nonnegative
integers such that $p+q\ge 2$. Denote by $x^\mu$ the $\mu$-th
standard coordinate for ${\bb R}^{p,q}$, and by $\eta$ the standard
indefinite metric tensor whose coordinate matrix $[\eta_{\mu\nu}]$
with respect to the standard basis of ${\bb R}^{p,q}$ is $\mr{diag}
(\underbrace{1, \cdots,1}_p, \underbrace{-1, \cdots -1}_q)$. As
usual, we use $[\eta^{\mu\nu}]$ to denote the inverse of
$[\eta_{\mu\nu}]$, $\mr{O}(p,q)$ to denote the set of endomorphisms
of ${\bb R}^{p,q}$ which preserve the quadratic form $\eta$, and
$\mr{O}^+(p,q)$ to denote the connected component of $\mr{O}(p,q)$
containing the identity. Note that it is customary to write
$\mr{O}(0, q)$ as $\mr{O}(q)$ and $\mr{O}(p,0)$ as $\mr{O}(p)$. The
followings are some basic topological facts about the
pseudo-orthogonal groups:

\begin{Fact}
1) $\mr{O}(p)$ is compact and has two connected components.

2) In the case both $p$ and $q$ are nonzero, $\mr{O}(p,q)$ is
non-compact and has four connected components. In fact, the
inclusion map $\mr{O}(p)\times \mr{O}(q)\to \mr{O}(p,q)$ is a
homotopic equivalence.

3) The inclusion map $\mr{O}^+(p)\times \mr{O}^+(q)\to
\mr{O}^+(p,q)$ is a homotopic equivalence. In fact, $\mr{O}^+(p)
\times \mr{O}^+(q)$ is a maximum compact subgroup of
$\mr{O}^+(p,q)$.

\end{Fact}

Let $\ms {C}_{p,q}$ be the Clifford algebra over $\bb C$ generated
by $X_\mu$'s subject to relations
$$
X_\mu X_\nu+X_\nu X_\mu=-2\eta_{\mu\nu}.
$$

Let $M_{\mu\nu}:={i\over 4}(X_\mu X_\nu-X_\nu X_\mu)$, then  one can
check that these $M$'s satisfy the following commutation relations:
\begin{eqnarray}
[M_{\alpha\beta},
M_{\gamma\delta}]=-i\left(\eta_{\beta\gamma}M_{\alpha\delta}
-\eta_{\alpha\gamma}M_{\beta\delta}
-\eta_{\beta\delta}M_{\alpha\gamma}
+\eta_{\alpha\delta}M_{\beta\gamma}\right).\nonumber
\end{eqnarray}

We use $\mr{Spin}(p)$ to denote the nontrivial double cover of
$\mr{SO}(p)$, and $\mr{Spin}(2,q)$ to denote the nontrivial double
cover of $\mr{O}^+(2,q)$ such that the inverse image of
$\mr{SO}(2)\times \mr{SO}(q)$ under the covering map is
$$\mr{Spin}(2)\times_{{\bb
Z}_2}\mr{Spin}(q):={\mr{Spin}(2)\times \mr{Spin}(q)\over{(g_1,
g_2)\sim (-g_1, -g_2)}}.$$ Note that $\mr{Spin}(2,q)$ defined here
is connected. We use $\widetilde{\mr{Spin}}(2,q)$ to denote the
unique double cover of $\mr{Spin}(2,q)$ such that the inverse image
of $\mr{Spin}(2)\times_{{\bb Z}_2}\mr{Spin}(q)$ under the covering
map is $\mr{Spin}(2)\times \mr{Spin}(q)$.

\subsection{Main Mathematical Results}
Let $G$ be one of the following real Lie groups: $\mr{Spin}(2n)$,
$\mr{Spin}(2n+1)$, $\mr{Spin}(2,2n)$,
$\widetilde{\mr{Spin}}(2,2n+1)$. We use $\frk{g}_0$ to denote the
Lie algebra of $G$ and $\frk g$ the complexification of $\frk{g}_0$.
In case $G$ is non-compact, we use $K$ to denote a maximal compact
subgroup of $G$.

When $G$ is compact, the representations of $G$ are all unitarizable
(hence reducible); moreover, an irreducible representation of $G$ is
precisely a finite dimensional highest weight modules of $\frk g$
with half integral weights.

When $G$ is non-compact, the (continuous) representations of $G$ are
not always unitarizable. It is known that a nontrivial unitarizable
module of $G$ must be infinite dimensional. By a fundamental theorem
of Harish-Chandra\footnote{See, for example, Theorem 7 on page 71 of
Ref. \cite{BK96}}, the irreducible unitary representations of $G$
are in one-one correspondence with the irreducible unitary $({\frk
g}, K)$-modules. Recall that a representation of $G$ is called a
highest weight representation if its underlying $({\frk g},
K)$-module is a highest weight $\frk g$-module. It is known from the
definitions and preceding quoted theorem of Harish-Chandra that a
highest weight representation of $G$ is an irreducible
representation of $G$. While the unitary highest weight
representations of $G$ has been classified in Refs.
\cite{Jakobsen81a,Jakobsen81b,EHW82}, a classification list for
unitary irreducible representations of $G$ is still missing in
general. Please note that a representation of $G$ is sometime also
called a $G$-module.

The following problem arises naturally from the construction and
analysis of the generalized hydrogen atoms.
\begin{Ques}\label{wProblem}
Classify all unitary highest weight representations of $G$ subject
to the following representation relations in the universal
enveloping algebra of $\frk{g}_0$: $\{M_{\mu\lambda},
{M^\lambda}_\nu\}=a\eta_{\mu\nu}$, i.e.,
\begin{eqnarray}\label{RepR}
\fbox{$M_{\mu\lambda}{M^\lambda}_\nu + {M^\lambda}_\nu
M_{\mu\lambda} =a\eta_{\mu\nu}$}
\end{eqnarray}
where $a$ is a representation-dependent real number, and
${M^\lambda}_\nu=\eta^{\lambda\delta}M_{\delta\nu}$.
\end{Ques}

It is not hard to see that $a$ is completely determined by the value
of the Casimir operator $c_2$ of $\frk{g}_0$ in a given
representation. In the case when $G$ is non-compact, Eq.
(\ref{RepR}) should be understood as an identity for operators on
the underlying $({\frk g}, K)$-module. Hereafter we shall call Eq.
(\ref{RepR}) the (quadratic) \underline{representation relations}.

\begin{Rmk} In the
compact case, the representation relations appear first in the
preliminary study of the dynamical symmetry of the generalized
MICZ-Kepler problems \cite{meng05}; and in the non-compact case, the
representation relations appear first in the study of MICZ-Kepler
problems \cite{Barut71}, and more recently in the refined study of
the dynamical symmetry of the generalized MICZ-Kepler problems
\cite{MZ07a, M07}.
\end{Rmk}

Throughout this paper, we adopt this practice in physics: the Lie
algebra generators act as hermitian operators in all unitary
representations.

The main mathematical results of this paper are summarized in the
following two theorems.

\begin{Th}[Compact Case]\label{RepTh1}  Let $n>0$ be an integer.

1) An irreducible unitary module of $\mr{Spin}(2n+1)$ satisfies Eq.
(\ref{RepR}) $\Leftrightarrow$ it is either the trivial
representation or the fundamental spin representation.

2) An irreducible unitary module of $\mr{Spin}(2n)$ satisfies Eq.
(\ref{RepR}) $\Leftrightarrow$ it is a Young power of a fundamental
spin representation.

\end{Th}

\begin{Th}[Non-Compact Case]\label{RepTh2}  Let $n>0$ be an integer.

1) A unitary highest weight module of
$\widetilde{\mr{Spin}}(2,2n+1)$ satisfies Eq. (\ref{RepR})
$\Leftrightarrow$ it is either the trivial one or the one with
highest weight\footnote{Unlike the case in part 2) of this theorem,
a representation here cannot descend to a representation of
$\mr{Spin}(2,2n+1)$.}
$$(-(n+\mu-{1\over 2}), \mu, \cdots, \mu)$$ for $\mu=0$ or $1/2$.

2) A unitary highest weight module of $\mr{Spin}(2,2n)$ satisfies
Eq. (\ref{RepR}) $\Leftrightarrow$ it is either the trivial one or
the one with highest weight $$(-(n+|\mu|-1), |\mu|, \cdots, |\mu|,
\mu)$$ for some half integer $\mu$.

\end{Th}
\begin{Rmk}
The representations characterized in part 1) are precisely the
Wallach representations in Case II ($\mu=0$) and Case III
($\mu=1/2$) on page 128 of Ref. \cite{EHW82}. The representations
characterized in part 2) are precisely the Wallach representations
in Case II ($\mu=0$), Case III ($\mu< 0$) and the mirror of Case III
($\mu>0$) on page 125 of Ref. \cite{EHW82}. In the
Enright-Howe-Wallach classification diagram for the unitary highest
weight modules, there are two reduction points in Case II and one
reduction point in Case III; the nontrivial representations
characterized here always sit on the first reduction point, and the
trivial representation (in Case II only) always sits on the 2nd
reduction point. In other word, the nontrivial representations
characterized here are precisely those boundary Wallach points in
Case II, Case III and the mirror of Case III (see, page 101, Ref.
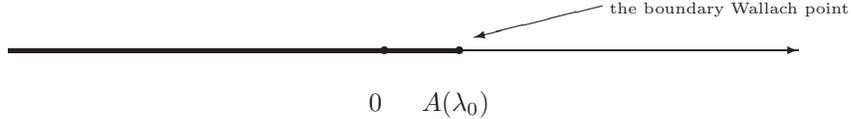
\cite{EHW82}):

\vskip 10pt \setlength{\unitlength}{1cm}
\begin{picture}(3,0.25)\linethickness{0.5mm}
\put(1,0){\line(1,0){6}} \put(6,0){\circle*{0.1}}\put(5.8,-0.8){$0$}
\put(1,0){\line(1,0){6}}
\put(7,0){\circle*{0.1}}\put(6.5,-0.8){$A(\lambda_0)$}
\put(9,0.5){\tiny {the boundary Wallach
point}}\put(8.9,0.6){\vector(-4,-1){1.7}} \thinlines
\put(1,0){\vector(1,0){10.5}}
\end{picture}
\end{Rmk}

\vskip 30pt The following subsection is about a corollary of Theorem
\ref{RepTh2} for the generalized hydrogen atoms and can be safely
ignored for readers who are only interested in mathematics.

\subsection{Main corollary for the generalized hydrogen atoms}
Let $D\ge 1$ be an integer, $\mu$ be a half integer if $D$ is even
and be $0$ or $1/2$ if $D$ is odd. To fix the terminology in this
paper, by the \emph{generalized hydrogen atom} in dimension $D$ with
magnetic charge $\mu$ we mean the hypothetic atom in dimension $D$
whose coulomb problem is the $D$-dimensional (quantum) MICZ-Kepler
problem with magnetic charge $\mu$ in the sense of Ref.
\cite{meng05}.

For the convenience of the readers, here we will give a quick review
of the $D$-dimensional (quantum) MICZ-Kepler problems. We assume
$D\ge 3$ and leave the case D = 1 or 2 to appendix \ref{Appendix1}.

Let $\bb R^D_*$ be the punctured $D$-space (i.e., $\bb R^D$ with the
origin removed), $\mr{S}^{D-1}$ be the unit sphere: $\{\vec r\mid
|\vec r|=1\}$. As we know, there is a canonical principal
$\mr{Spin}(D-1)$-bundle $\mr{Spin}(D)\to \mr{S}^{D-1}$ with a
canonical connection\footnote{The connection form is
$\mr{Pr}_{\frk{so}(D-1)}(g^{-1}\,dg)$, where $g^{-1}\,dg$ is the
Maurer-Cartan form and $\mr{Pr}_{\frk{so}(D-1)}$ is the orthogonal
projection from $\frk{so}(D)$ onto $\frk{so}(D-1)$.}. Via the
natural retraction map $\bb R^D_*\to \mr{S}^{D-1}$, we get a
canonical principal $\mr{Spin}(D- 1)$-bundle with a canonical
connection over $\bb R^D_*$. By choosing the representation of
$\frk{so}(D-1)$ with highest weight $(|\mu|,\cdots,|\mu|,\mu)$, we
get an associated hermitian vector bundle with an hermitian
connection on Riemannian manifold $({\bb R}^D_*; dx_1^2 + \cdots+
dx_D^2)$.

This bundle is denoted by ${\mathcal S}^{2\mu}$ and it is our
analogue of the Dirac monopole with magnetic charge $\mu$. By
definition, the {\em $D$-dimensional MICZ-Kepler problem with
magnetic charge $\mu$} is defined to be the quantum mechanical
system on ${\bb R}_*^D$ for which the wave-functions are sections of
${\mathcal S}^{2\mu}$ and the hamiltonian is
\begin{eqnarray}
H=\left\{
\begin{array}{rl}
-{1\over 2}\Delta_\mu+{\mu^2+(n-1)|\mu|\over 2r^2}-{1\over r} &
\hbox{if $D=2n+1$}\\
\\
-{1\over 2}\Delta_\mu+{(n-1)\mu\over 2r^2}-{1\over r} & \hbox{if
$D=2n$}
\end{array}\right.
\end{eqnarray}
where $\Delta_\mu$ is the standard Laplace operator
$\partial_1^2+\cdots+\partial_D^2$ twisted by ${\mathcal S}^{2\mu}$.

Physically it is interesting to find all square integrable
eigen-sections of $H$. It has been shown in Refs. \cite{MZ07a, M07}
that the linear span of the square integrable eigen-sections of H is
a unitary highest weight Harish-Chandra module with highest weight
$$(-\left({D-1\over 2}+|\mu|\right),|\mu|,\cdots,|\mu|,\mu).$$
Recall that, the Hilbert space completion of this linear span is
called {\em the Hilbert space of bound states of the $D$-dimensional
generalized hydrogen atom with magnetic charge $\mu$}; so, in view
of the fundamental theorem of Harish-Chandra we quoted earlier, it
is a nontrivial unitary highest weight representation of
$\widetilde{\mr{Spin}}(2,D + 1)$.

It has been shown in Refs. \cite{MZ07a, M07} that such a unitary
highest weight representation of $\widetilde{\mr{Spin}}(2,D + 1)$
has a very explicit geometric realization. To describe it, we let
$d^Dx$ be the Lebesgue measure on ${\bb R}^D$. The Hilbert space of
square integrable (with respect to $d^Dx$) sections of ${\mathcal
S}^{2\mu}$ (denoted by $L^2({\mathcal S}^{2\mu})$), being identified
with {\em the twisted Hilbert space of bound states of the
$D$-dimensional generalized hydrogen atom with magnetic charge
$\mu$}, turns out to be the representation space. To describe the
unitary action of $\widetilde{\mr{Spin}}(2, D+1)$ on $L^2({\mathcal
S}^{2\mu})$, we just need to describe the infinitesimal action on
$C^\infty({\mathcal S}^{2\mu})$; for that purpose, it suffices to
describe how $M_{\alpha,0}$ ($1\le \alpha\le D$), $M_{D+1,0}$ and
$M_{-1, 0}$ act as differential operators: they act as $i\sqrt
r\nabla_\alpha\sqrt r$, ${1\over 2}\left(\sqrt r\Delta_\mu\sqrt
r+r-{c\over r}\right)$ and ${1\over 2}\left(\sqrt r\Delta_\mu\sqrt
r-r-{c\over r}\right)$ respectively.  For example, for $\psi\in
C^\infty({\mathcal S}^{2\mu})$, we have
\begin{eqnarray}
(M_{\alpha,0}\cdot \psi)(r, \Omega) & = & i\sqrt r\nabla_\alpha
\left (\sqrt r\psi(r, \Omega)\right).\nonumber
\end{eqnarray}

Therefore, together with the results in appendix \ref{Appendix1}, we
have the following corollary of Theorem \ref{RepTh2} for the
generalized hydrogen atoms.

\begin{Th}[Main Theorem] Let $D\ge 1$ be an integer.

1) The Hilbert space of bound states of a $D$-dimensional
generalized hydrogen atom always forms a nontrivial unitary highest
weight representation of $\widetilde{\mr{Spin}}(2, D+~1)$.

2) A nontrivial unitary highest weight representation of
$\widetilde{\mr{Spin}}(2, D+1)$ can be realized by the Hilbert space
of bound states of a $D$-dimensional generalized hydrogen atom
$\Leftrightarrow$ it satisfies the quadratic representation
relations.
\end{Th}
Therefore, the Hilbert spaces of bound states for $D$-dimensional
generalized hydrogen atoms realize precisely the nontrivial Wallach
points for $\widetilde{\mr{Spin}}(2, D+1)$ listed in Case II, Case
III and the mirror of Case III on page 127 (when $D$ is odd), and on
page 125 (when $D$ is even) in Ref. \cite{EHW82}. Note that when $D$
is odd, the mirror of Case III is Case $I_p$ with $p={D+1\over 2}$;
when $D$ is odd, Case III = the mirror of Case III.

We end this subsection with the following concluding remark.
\begin{Rmk}
The interesting families of representations of
$\widetilde{\mr{Spin}}(2, D+1)$ form the following descending chain:

{\em \{admissible irreps\} $\supset$ \{unirreps\} $\supset$ \{H.WT.
unitary reps\}  $\supset$ \{Wallach reps\}  $\supset$ \{nontrivial
Wallach reps of type II, type III or mirror of type III\}. }

For the bottom family of representations in this chain, combining
the results from Refs. \cite{MZ07a, M07}, one can reach the
following conclusions:

{\em

1) The members of this family can be precisely realized as the
Hilbert space of bound states for generalized hydrogen atoms in
dimension $D$;

2) Each member of this family can be realized as the Hilbert space
of $L^2$-sections of a canonical hermitian bundle over ${\bb R}^D_*$
equipped with a canonical hermitian connection;

3) This family can be characterized by a canonical finite set of
quadratic relations among the infinitesimal generators of
$\widetilde{\mr{Spin}}(2, D+1)$.}

\end{Rmk}

\subsection{Outline of the paper}
As a warm up, we will first give a proof of Theorem \ref{RepTh1} in
section \ref{STh1}, the idea is essentially taken from the appendix
of Ref. \cite{meng05} and the arguments are purely algebraic. Then
we prove Theorem \ref{RepTh2} by similar arguments in section
\ref{STh2}. I would like to thank Qi You for simplifying the proof
of part 2) of Theorem \ref{RepTh1}.

\section{Proof of Theorem \ref{RepTh1}}\label{STh1}
We will follow the approach in the appendix of Ref. \cite{meng05}.
The idea is to find a convenient Cartan basis and then rewrite the
representation relations in terms of these Cartan basis elements. We
start with the proof of part 1) because it is technically simpler.
The proof of part 2) is similar, but technically is a bit more
involved.

\subsection{Part 1)} We assume that $n\ge 1$. To continue, a
digression on Lie algebra $\frk{so}(2n+1)$ is needed. Recall that
the root space of $\frk{so}(2n+1)$ is $\bb R^n$. Let $e^i$ be the
vector in $\bb R^n$ whose $i$-th entry is $1$ and all other entries
are zero. The positive roots are $e^i\pm e^j$ with $1\le i<j\le n$
and $e^k$ with $1\le k\le n$. Following Ref. \cite{georgi82}, we
choose the following Cartan basis for $\frk{so}(2n+1)$:
\begin{eqnarray}
\left\{ \begin{array}{rcl} H_i & = & M_{2i-1, 2i}\quad\quad
\hbox{$1\le i\le n$}\cr E_{\eta e^j+\eta' e^k} & = & {1\over
2}\left(M_{2j-1, 2k-1}+i\eta M_{2j, 2k-1}+i\eta'M_{2j-1,
2k}-\eta\eta'M_{2j,2k}\right)\cr &&\quad\quad\mbox{for $j< k$}\cr
E_{\eta e^j} & = & {1\over \sqrt 2}\left(M_{2j-1, 2n+1}+i\eta M_{2j,
2n+1}\right)\;\mbox{for $j\le n$}\end{array}\right.\nonumber
\end{eqnarray} where $\eta, \eta'\in\{1, -1\}$. For convenience, we
also use the same expression above to define $E_{\eta e^j+\eta'
e^k}$ when $j>k$, then we have $$E_{\eta e^j+\eta' e^k}=-E_{\eta'
e^k+\eta e^j}$$ for $j\neq k$.

We are interested in unitary representations, i.e., representations
such that each $M_{ij}$ acts as an hermitian operator, or
equivalently, each $H_i$ act as an hermitian operator, and
$$
(E_{\alpha})^\dag=E_{-\alpha}.
$$

Let $|\Omega\rangle=|\lambda_1\cdots\lambda_n\rangle$ be the highest
weight state of a unitary representation for which the
representation relations hold. So $H_i|\Omega\rangle=\lambda_i
|\Omega\rangle$ and $E_\alpha|\Omega\rangle=0$ if $\alpha$ is a
positive root. Since
\begin{eqnarray}
[E_{e^i\pm e^j}, E_{-e^i\mp e^j}]=H_i\pm H_j,\quad [E_{e^i},
E_{-e^i}]=H_i,
\end{eqnarray}by the unitarity, we conclude that \begin{eqnarray}
\lambda_1\ge\cdots\ge\lambda_n\ge 0.\end{eqnarray} Since $\{E_{e^i},
E_{-e^i}, H_i\}$ span the Lie algebra of $\frk{su}(2)$, and the
orbit of $|\Omega\rangle$ under the action of the universal
enveloping algebra of this $\frk{su}(2)$ is a highest weight
representation with $|\Omega\rangle$ as its highest weight state, we
conclude that $\lambda_i$ is a half integer. A similar argument
shows that $\lambda_i-\lambda_j$ is always an integer.

\vskip 10pt \noindent $\Rightarrow$:  The representation relations
say that, for $1\le j\le n+1$, we have
\begin{eqnarray}
\langle \Omega |\sum_k(M_{2j-1,k})^2 |\Omega\rangle =c,
\end{eqnarray} where $c$ is a constant independent of $j$.
Since
\begin{eqnarray}\left\{ \begin{array}{rcl}
\sum_{k}(M_{2j-1,k})^2& =&H_j^2+{H_j\over 2} +{1\over 2}\sum_{i\neq
j}
\left(\{E_{-e^j-e^i},E_{e^j+e^i}\}+\{E_{-e^j+e^i},E_{e^j-e^i}\}\right)\cr
& & +{1\over 2}((E_{-e^j})^2+(E_{e^j})^2)+E_{-e^j}E_{e^j}\cr & &+
\sum_{i\neq
j}\left(E_{-e^j-e^i}E_{-e^j+e^i}+E_{e^j+e^i}E_{e^j-e^i}\right)\quad\mbox{for
$1\le j\le n$},\cr \sum_k (M_{2n+1,k})^2& =& \sum_{i}\{E_{e^i},
E_{-e^i}\},\end{array}\right.\nonumber
\end{eqnarray}
We have
\begin{eqnarray}\label{identitiesParta}\left\{ \begin{array}{rcl}
\lambda_1^2+(n-{1\over 2})\lambda_1 &= &c\cr \lambda_2^2+(n-{1\over
2})\lambda_2+(\lambda_1-\lambda_2) & = & c\cr \lambda_3^2+(n-{1\over
2})\lambda_3+(\lambda_1+\lambda_2-2\lambda_3) & = & c\cr &\vdots
&\cr \lambda_n^2+(n-{1\over
2})\lambda_n+(\lambda_1+\cdots+\lambda_{n-1}-(n-1)\lambda_n) & = & c
\cr \sum\lambda_i &=& c.\end{array}\right.
\end{eqnarray}
Subtracting 2nd identity from the 1st identity, we have
$$
(\lambda_1-\lambda_2)(\lambda_1+\lambda_2+n-{3\over 2})=0.
$$So $\lambda_1=\lambda_2=\lambda$ if
$n\ge 2$. Assume $n\ge 3$, subtracting the 3rd identity from the 1st
identity, we have
$$
(\lambda-\lambda_3)(\lambda+\lambda_2+n-{5\over 2})=0.
$$ So $\lambda_3=\lambda$ if
$n\ge 3$. By repeating this argument $(n-1)$ times, we get
$\lambda_1=\cdots=\lambda_n=\lambda$. Then
$|\Omega\rangle=|\lambda\cdots\lambda\rangle$.

By equating the 1st identity with the last identity, we have
$$
\lambda^2={1\over 2}\lambda,
$$then $\lambda=0$ or $1/2$. The case
that $\lambda=0$ corresponds to the trivial representation and the
case that $\lambda =1/2$ corresponds to the fundamental spin
representation.

\vskip 10pt \noindent $\Leftarrow$:  The representation relations
are trivially true in the former case, and can be checked easily by
using Clifford algebra in the later case: $M_{jk}\propto e_je_k$, so
$$\{M_{jk}, M_{kl}\}\propto
e_je_ke_ke_l+e_ke_le_je_k\propto -e_je_l-e_le_j=2\delta_{jl}.$$

End of the proof of part 1) of Theorem \ref{RepTh1}.

\subsection{Part 2)} It is trivial
when $n=1$. So we assume that $n\ge 2$. To continue, a digression on
Lie algebra $\frk{so}(2n)$ is needed. Recall that the root space of
$\frk{so}(2n)$ is $\bb R^n$. Let $e^i$ be the vector in $\bb R^n$
whose $i$-th entry is $1$ and all other entries are zero. The
positive roots are $e^i\pm e^j$ with $1\le i<j\le n$. Following Ref.
\cite{georgi82}, we choose the following Cartan basis for
$\frk{so}(2n)$:
\begin{eqnarray}\left\{ \begin{array}{rcl} H_i & = & M_{2i-1, 2i}\quad\quad \hbox{$1\le i\le
n$}\cr E_{\eta e^j+\eta' e^k} & = & {1\over 2}\left(M_{2j-1,
2k-1}+i\eta M_{2j, 2k-1}+i\eta'M_{2j-1,
2k}-\eta\eta'M_{2j,2k}\right)\cr & & \quad\quad\quad\mbox{for $j<
k$}\end{array}\right.\nonumber
\end{eqnarray} where $\eta, \eta'\in\{1, -1\}$. For convenience, we
also use the same expression above to define $E_{\eta e^j+\eta'
e^k}$ for $j>k$, then we have $$E_{\eta e^j+\eta' e^k}=-E_{\eta'
e^k+\eta e^j}$$ for $j\neq k$.

Let $|\Omega\rangle=|\lambda_1\cdots\lambda_n\rangle$ be the highest
weight state of a unitary representation for which the
representation relations hold. So $H_i|\Omega\rangle=\lambda_i
|\Omega\rangle$ and $E_\alpha|\Omega\rangle=0$ if $\alpha$ is a
positive root.

Since
\begin{eqnarray}
[E_{e^i\pm e^j}, E_{-e^i\mp e^j}]=H_i\pm H_j,
\end{eqnarray}by the unitarity, we conclude that \begin{eqnarray}
\lambda_1\ge\cdots\ge\lambda_{n-1}\ge |\lambda_n|.\end{eqnarray}
Since $\{E_{e^i+e^j}, E_{-e^i-e^j},{1\over 2}(H_i+H_j)\}$ span the
Lie algebra of $\frk{su}(2)$, we conclude that $\lambda_i-\lambda_j$
 is an integer. A similar argument shows that $\lambda_i-\lambda_j$
is an integer. So $\lambda_i$'s are half integers.

\vskip 10pt\noindent $\Rightarrow$:  The representation relations
say that, for $1\le j\le n$, we have
\begin{eqnarray}\label{1stRepRe}
\langle \Omega |\sum_k(M_{2j-1,k})^2 |\Omega\rangle =c,
\end{eqnarray} where $c$ is a constant independent of $j$.
Since $$ \sum_k(M_{2j-1,k})^2 = H_j^2+{1\over 2}\sum_{i\neq j}
\left(\{E_{-e^j-e^i},E_{e^j+e^i}\}+\{E_{-e^j+e^i},E_{e^j-e^i}\}\right),
$$
we have
\begin{eqnarray}\left\{ \begin{array}{rcl}
\lambda_1^2+(n-1)\lambda_1 &= &c\cr
\lambda_2^2+(n-1)\lambda_2+(\lambda_1-\lambda_2) & = & c\cr
\lambda_3^2+(n-1)\lambda_3+(\lambda_1+\lambda_2-2\lambda_3) & = &
c\cr &\vdots &\cr
\lambda_n^2+(n-1)\lambda_n+(\lambda_1+\cdots+\lambda_{n-1}-(n-1)\lambda_n)
& = & c.\end{array}\right.
\end{eqnarray}
Subtracting the 2nd identity from the 1st identity, we have
$$
(\lambda_1-\lambda_2)(\lambda_1+\lambda_2+n-2)=0.
$$So if $\lambda_1=|\lambda_2|$ if $n=2$, and $\lambda_1=\lambda_2=\lambda$ if
$n>2$. Assume $n\ge 3$, subtracting the 3rd identity from the 1st
identity, we have
$$
(\lambda-\lambda_3)(\lambda+\lambda_2+n-3)=0.
$$ So if $\lambda=|\lambda_3|$ if $n=3$, and $\lambda_3=\lambda$ if
$n>3$. By repeating this argument $(n-1)$ times, we get
$\lambda_1=\cdots=\lambda_{n-1}=\lambda$ and $\lambda=|\lambda_n|$.
Then the representation must be a Young power of a fundamental spin
representation.

\vskip 10pt\noindent $\Leftarrow$: We need to prove that the
representation relations (i.e., Eq. (\ref{RepR})) hold for any Young
power of a fundamental spin representation. The proof is broken into
three steps, with the last one being significantly simplified by Qi
You.

\underline{Step one}. We may assume the representation is $\mb
s_+^{2\mu}$ for some non-negative half integer $\mu$. That is
because there exists a $g\in \mbox{Pin}(2n)$ such that the action by
$g$ on $\mb s_-^{2\mu}\oplus \mb s_+^{2\mu}$ produces a vector space
isomorphism: $\mb s_-^{2\mu}\to \mb s_+^{2\mu}$, moreover,
$gM_{1,k}g^{-1}=-M_{1,k}$ and $gM_{j,k}g^{-1}=M_{j,k}$ for $1<j<k$;
consequently, the representation relations are invariant under the
(adjoint) action by $g$.

\underline{Step two}. For any $i<j$,  relation \begin{eqnarray}
\sum_k\{M_{i,k},M_{j, k}\} = 0
\end{eqnarray} hold for $\mb
s_+^{2\mu}$.
\begin{proof} It suffice to prove the statement in the case $i=1$
and $j=2$; that is because, for any $i'<j'$, there is an element in
$g\in \mbox{Spin}(2n)$ such that $$g\sum_k\{M_{1,k},M_{2,
k}\}g^{-1}=\sum_k\{M_{i',k},M_{j', k}\}.$$

Next we observe that $$ \sum_k\{M_{1,k},M_{2, k}\} = {2\over
i}(\ms{O}^\dag-\ms{O}) $$ where
$$\ms{O}=\sum_{i\neq 1}E_{-e^1-e^i}E_{-e^1+e^i}.$$
Consequently, we can finish the proof by showing that
\begin{eqnarray}\label{olambda=0}\ms{O}|\Lambda\rangle=0\end{eqnarray} for any $|\Lambda\rangle \in \mb
s_+^{2\mu}$. But that is OK because of the following easy facts:
\begin{eqnarray}[\ms{O}, E_{-\alpha}]&=&0 \quad \mbox{for any positive root $\alpha$}, \cr
E_{-e^1+e^i}|\Omega\rangle & = & 0\quad\mbox{where $|\Omega\rangle=
|\underbrace{\mu\cdots\mu}_n\rangle$,}\nonumber
\end{eqnarray} and the fact that $|\Lambda\rangle$ is a linear
combination of the states created from $|\Omega\rangle$ by some
$E_{-\alpha}$'s with $\alpha$ being positive roots.
\end{proof}

\underline{Step three}. For any $j$,  relation
\begin{eqnarray}
\sum_k(M_{j,k})^2 - {1\over n}c_2 &=& 0
\end{eqnarray} hold for $\mb
s_+^{2\mu}$. In fact, it suffices to show that relation
\begin{eqnarray}
\sum_k(M_{1,k})^2 - {1\over n}c_2 &=& 0
\end{eqnarray} hold for $\mb
s_+^{2\mu}$.

\begin{proof}
Observe that\footnote{The much simplified proof presented here is
due to this key observation by Qi You}
\begin{eqnarray}
[M_{ab}, \sum_k(M_{1,k})^2 - {1\over n}c_2] &=&
-i\eta_{b1}\sum_k\{M_{ak}, M_{1k}\}+i\eta_{a1}\sum_k\{M_{bk},
M_{1k}\}\cr &=& 0\quad \mbox{on $\mb s_+^{2\mu}$ by step two above.
}\nonumber
\end{eqnarray}
Therefore, it suffices to show that
\begin{eqnarray}
\left( \sum_k(M_{1,k})^2 - {1\over n}c_2 \right) |\Omega\rangle=0.
\end{eqnarray}
But that is not hard, because
\begin{eqnarray}\sum_k(M_{1,k})^2 - {1\over n}c_2 & = &
\ms{O}_1+\ms{O}^\dag+\ms{O}\cr &=& \ms{O}_1 \quad \mbox{on $\mb
s_+^{2\mu}$ by Eq. (\ref{olambda=0}). }\cr &=&0\quad \mbox{on
$|\Omega\rangle$ by a straight forward calculation, }\nonumber
\end{eqnarray} where
\begin{eqnarray}\ms{O}_1 &= &
H_1^2-{c_2\over n}+{1\over 2}\sum_{i\neq 1}
\left(\{E_{-e^1-e^i},E_{e^1+e^i}\}+\{E_{-e^1+e^i},E_{e^1-e^i}\}\right).
\nonumber
\end{eqnarray}
\end{proof}
Steps two and three together say that the representation relations
hold in $\mb s_+^{2\mu}$, hence also hold in $\mb s_-^{2\mu}$ by
step one.

End of the proof of part 2) of Theorem \ref{RepTh1}.

\section{Proof of Theorem \ref{RepTh2}}\label{STh2}
The proof of theorem \ref{RepTh2} is similar to that of theorem
\ref{RepTh1}, but technically more involved. Again, we start with
the proof of part 1). Although a straightforward proof does exist,
to make the proof shorter, we use results from both Refs.
\cite{MZ07a, M07} and appendix \ref{Appendix1}.

\subsection{Part 1)} We assume that $n\ge 1$. To continue, a
digression on Lie algebra $\frk{so}(2,2n+1)$ is needed. Recall that
the root space of $\frk{so}(2,2n+1)$ is $\bb R^{n+1}$. Let $e^i$ be
the vector in $\bb R^{n+1}$ whose $i$-th entry is $1$ and all other
entries are zero. The positive roots are $e^i\pm e^j$ with $0\le
i<j\le n$ and $e^k$ with $0\le k\le n$.

Following Ref. \cite{georgi82}, we choose the following Cartan basis
for $\frk{so}(2, 2n+1)$:
\begin{eqnarray}\left\{ \begin{array}{rcl} H_0 &= &M_{-1, 0},\cr H_i & = & -M_{2i-1, 2i}\quad\quad \hbox{$1\le i\le
n$},\cr E_{\eta e^j+\eta' e^k} & = & {1\over 2}\left(M_{2j-1,
2k-1}+i\eta M_{2j, 2k-1}+i\eta'M_{2j-1,
2k}-\eta\eta'M_{2j,2k}\right)\cr & & \quad\quad\quad\mbox{for $0\le
j< k\le n$},\cr E_{\eta e^j} & = & {1\over \sqrt 2}\left(M_{2j-1,
2n+1}+i\eta M_{2j, 2n+1}\right)\quad \mbox{for $0\le j\le n$},
\end{array}\right.\nonumber
\end{eqnarray} where $\eta, \eta'\in\{1, -1\}$. For convenience, we
also use the same expression above to define $E_{\eta e^j+\eta'
e^k}$ for $j>k$, then we have $$E_{\eta e^j+\eta' e^k}=-E_{\eta'
e^k+\eta e^j}$$ for $j\neq k$.

Let $|\Omega\rangle=|\lambda_0\lambda_1\cdots\lambda_n\rangle$ be
the highest weigh state of a unitary representation for which the
representation relations hold. So $H_i|\Omega\rangle=\lambda_i
|\Omega\rangle$ and $E_\alpha|\Omega\rangle=0$ if $\alpha$ is a
positive root.

Since
\begin{eqnarray}\left\{ \begin{array}{rcl}
[E_{e^0+e^i}, E_{-e^0-e^i}]=-H_0-H_i,\quad [E_{e^0-e^i},
E_{-e^0+e^i}]=-H_0+H_i,\cr [E_{e^i+e^j}, E_{-e^i-e^j}]=H_i+H_j,\quad
[E_{e^i-e^j}, E_{-e^i+e^j}]=H_i-H_j,\cr [E_{\eta e^i}, E_{\eta'
e^j}]=-iE_{\eta e^i+\eta'e^j},\; [E_{e^i}, E_{-e^i}]=H_i,\quad
[E_{e^0}, E_{-e^0}]=-H_0,
\end{array}\right.\nonumber
\end{eqnarray} by unitarity, we conclude that \begin{eqnarray}-\lambda_0\ge
\lambda_1\ge\cdots\ge\lambda_n\ge 0.\end{eqnarray} For $i\neq 0$,
$\{E_{e^i}, E_{-e^i}, H_i\}$ span the Lie algebra of $\frk{su}(2)$,
then $\lambda_i$ must be a half integer. A similar argument shows
that $\lambda_i-\lambda_j$ is an integer for $0<i<j\le n$.

\vskip 10pt \noindent $\Rightarrow$:  The representation relations
say that
\begin{eqnarray}\left\{ \begin{array}{rcl}
\langle \Omega |-\sum M_{-1,k}{M^k}_{-1} |\Omega\rangle  &  = & c,
\cr \\\langle \Omega |\sum M_{2j-1,k}{M^k}_{2j-1} |\Omega\rangle  &
= & c\quad\mbox{for $j=1, 2, \cdots, n+1$},
\end{array}\right.
\end{eqnarray} where $c$ is a constant.
Since
\begin{eqnarray}\left\{ \begin{array}{rcl}\sum M_{-1,k}{M^k}_{-1}& =&-H_0^2-{H_0\over 2}
+{1\over 2}\sum_{i\neq 0}
\left(\{E_{-e^0-e^i},E_{e^0+e^i}\}+\{E_{-e^0+e^i},E_{e^0-e^i}\}\right)\cr
& & +{1\over 2}((E_{-e^0})^2+(E_{e^0})^2)+E_{-e^0}E_{e^0}\cr & &+
\sum_{i\neq
0}\left(E_{-e^0-e^i}E_{-e^0+e^i}+E_{e^0+e^i}E_{e^0-e^i}\right), \cr
\\
\sum M_{2j-1,k}{M^k}_{2j-1}& =&H_j^2+{H_j\over 2} +{1\over
2}\sum_{i\neq 0, j}
\left(\{E_{-e^j-e^i},E_{e^j+e^i}\}+\{E_{-e^j+e^i},E_{e^j-e^i}\}\right)\cr
& & +{1\over 2}((E_{-e^j})^2+(E_{e^j})^2)+E_{-e^j}E_{e^j}\cr & &+
\sum_{i\neq 0,
j}\left(E_{-e^j-e^i}E_{-e^j+e^i}+E_{e^j+e^i}E_{e^j-e^i}\right)\cr &
& -{1\over 2}
\left(\{E_{-e^0-e^j},E_{e^0+e^j}\}+\{E_{-e^0+e^j},E_{e^0-e^j}\}\right)\cr
& & -\left(E_{-e^0-e^j}E_{-e^0+e^j}+E_{e^0+e^j}E_{e^0-e^j}\right),
\cr
\\
\sum M_{2n+1,k}{M^k}_{2n+1}& =&-\{E_{e^0},
E_{-e^0}\}+\sum_{i>0}\{E_{e^i}, E_{-e^i}\},
\end{array}\right.\nonumber
\end{eqnarray}
we have
\begin{eqnarray}\label{identitiesParta}
\left\{ \begin{array}{rcl} \lambda_0^2+(n+{1\over 2})\lambda_0 &=
&c\cr \lambda_1^2+(n-{1\over 2})\lambda_1+\lambda_0 & = & c\cr
\lambda_2^2+(n-{1\over 2})\lambda_2+(\lambda_1-\lambda_2)+\lambda_0
& = & c \cr &\vdots &\cr \lambda_n^2+(n-{1\over
2})\lambda_n+(\lambda_1+\cdots+\lambda_{n-1}-(n-1)\lambda_n)+\lambda_0
& = & c\cr \sum\lambda_i & =& c.
\end{array}\right.
\end{eqnarray}
Subtracting the 3rd identity from the 2nd identity, we have
$$
(\lambda_1-\lambda_2)(\lambda_1+\lambda_2+n-{3\over 2})=0.
$$So $\lambda_1=\lambda_2=\lambda$ if
$n\ge 2$. Assume $n\ge 3$, subtracting the 4th identity from the 3rd
identity, we have
$$
(\lambda-\lambda_3)(\lambda+\lambda_2+n-{5\over 2})=0.
$$ So $\lambda_3=\lambda$ if
$n\ge 3$. By repeating this argument $(n-1)$ times, we get
$\lambda_1=\cdots=\lambda_n=\lambda$. Then
$|\Omega\rangle=|\lambda_0\lambda\cdots\lambda\rangle$.

By comparing the 2nd with the last identities, we get
$$
\lambda^2={1\over 2}\lambda,
$$ so $\lambda=0$ or $1/2$.

By comparing the first two identities, we get
$$
(\lambda-\lambda_0)(\lambda+\lambda_0+n-{1\over 2})=0.
$$
So either $\lambda_0=\lambda$ or $\lambda_0=-(\lambda+n-{1\over
2})$. In view of the fact that $-\lambda_0\ge \lambda$, we conclude
that $(\lambda_0, \lambda)$ must be one of the following three
pairs: $(0,0)$, $(-n+1/2, 0)$, $(-n, 1/2)$. Consequently, the
unitary highest weight representation, if it exists, must be one of
the following three cases: 1) the trivial one, 2) the one with
highest weight $(-n+1/2, 0,\ldots, 0)$, 3) the one with highest
weight $(-n, 1/2,\ldots, 1/2)$.

\vskip 10pt\noindent $\Leftarrow$:  The remaining question we must
answer is this: such representations do exist and satisfy the
representation relations. This is certainly clear in the trivial
case.

The existence of such representations in the nontrivial case is
clear from the classification result of Refs. \cite{Jakobsen81a,
Jakobsen81b,EHW82}. As a matter of fact, in view of Theorem 1 in
Ref. \cite{M07}, the one with highest weight $(-n+1/2, 0,\ldots, 0)$
can be realized by the $2n$-dimensional generalized Kepler problem
with magnetic charge $0$, and the one with highest weight $(-n,
1/2,\ldots, 1/2)$ can be realized by the $2n$-dimensional
generalized Kepler problem with magnetic charge $1/2$. Moreover, in
view of part 2) of Theorem 2 in Ref. \cite{M07}, these
representations indeed satisfy the representation relations. The
only problem with this argument is that the case $n=1$ is not
covered; however, the results in Refs. \cite{meng05, M07} can be
pushed down to the case $n=1$, see subsection \ref{micz2} in
appendix \ref{Appendix1}.

End of the proof of part 1) of Theorem \ref{RepTh2}.

\vskip 10pt We would like to remark that, by following the argument
in the proof of part 2) of Theorem \ref{RepTh1}, one can also verify
the representation relations directly. Since this argument is a bit
long, we choose to skip it.

\subsection{Part 2)}
We assume that $n\ge 1$. To continue, a digression on Lie algebra
$\frk{so}(2,2n)$ is needed. Recall that the root space of
$\frk{so}(2,2n)$ is $\bb R^{n+1}$. Let $e^i$ be the vector in $\bb
R^{n+1}$ whose $i$-th entry is $1$ and all other entries are zero.
The positive roots are $e^i\pm e^j$ with $0\le i<j\le n$.

Following Ref. \cite{georgi82}, we choose the following Cartan basis
for $\frk{so}(2, 2n)$:
\begin{eqnarray}\left\{ \begin{array}{rcl} H_0 &= &M_{-1, 0},\cr H_i & = & -M_{2i-1, 2i}\quad\quad \hbox{$1\le i\le
n$},\cr E_{\eta e^j+\eta' e^k} & = & {1\over 2}\left(M_{2j-1,
2k-1}+i\eta M_{2j, 2k-1}+i\eta'M_{2j-1,
2k}-\eta\eta'M_{2j,2k}\right)\cr & & \quad\quad\quad\mbox{for $0\le
j< k\le n$}.\end{array}\right.\nonumber
\end{eqnarray} Here $\eta, \eta'\in\{1, -1\}$. For convenience, we
also use the same expression above to define $E_{\eta e^j+\eta'
e^k}$ for $j>k$, then we have $$E_{\eta e^j+\eta' e^k}=-E_{\eta'
e^k+\eta e^j}$$ for $j\neq k$.

Let $|\Omega\rangle=|\lambda_0\lambda_1\cdots\lambda_n\rangle$ be
the highest weigh state of a representation for which the
representation relations hold. So $H_i|\Omega\rangle=\lambda_i
|\Omega\rangle$ and $E_\alpha|\Omega\rangle=0$ if $\alpha$ is a
positive root.

Since
\begin{eqnarray}\left\{ \begin{array}{rcl}
[E_{e^0+e^i}, E_{-e^0-e^i}]=-H_0-H_i,\quad [E_{e^0-e^i},
E_{-e^0+e^i}]=-H_0+H_i,\cr [E_{e^i+e^j}, E_{-e^i-e^j}]=H_i+H_j,\quad
[E_{e^i-e^j}, E_{-e^i+e^j}]=H_i-H_j,\end{array}\right.\nonumber
\end{eqnarray} by unitarity, we conclude that \begin{eqnarray}-\lambda_0\ge
\lambda_1\ge\cdots\ge\lambda_{n-1}\ge |\lambda_n|.\end{eqnarray}
Just as before, one can show that each $\lambda_i$ with $i>0$ is a
half integer and each $\lambda_i-\lambda_j$ with $0<i<j\le n$ is an
integer.

\vskip 10pt \noindent $\Rightarrow$:  The representation relations
say that
\begin{eqnarray}\left\{ \begin{array}{rcl}
\langle \Omega |-\sum M_{-1,k}{M^k}_{-1}|\Omega\rangle  &  = & c,
\cr \langle \Omega |\sum M_{2j-1,k}{M^k}_{2j-1} |\Omega\rangle  &  =
& c\quad\mbox{for $j=1, 2, \cdots, n$},\end{array}\right.
\end{eqnarray} where $c$ is a constant. Since
\begin{eqnarray}\left\{ \begin{array}{rcl}\sum M_{-1,k}{M^k}_{-1}& =&-H_0^2
+{1\over 2}\sum_{i\neq 0}
\left(\{E_{-e^0-e^i},E_{e^0+e^i}\}+\{E_{-e^0+e^i},E_{e^0-e^i}\}\right)\cr
& &+ \sum_{i\neq
0}\left(E_{-e^0-e^i}E_{-e^0+e^i}+E_{e^0+e^i}E_{e^0-e^i}\right),
\cr\\
\sum M_{2j-1,k}{M^k}_{2j-1}& =&H_j^2 +{1\over 2}\sum_{i\neq 0, j}
\left(\{E_{-e^j-e^i},E_{e^j+e^i}\}+\{E_{-e^j+e^i},E_{e^j-e^i}\}\right)\cr
& &+ \sum_{i\neq 0,
j}\left(E_{-e^j-e^i}E_{-e^j+e^i}+E_{e^j+e^i}E_{e^j-e^i}\right)\cr &
& -{1\over 2}
\left(\{E_{-e^0-e^j},E_{e^0+e^j}\}+\{E_{-e^0+e^j},E_{e^0-e^j}\}\right)\cr
& &
-\left(E_{-e^0-e^j}E_{-e^0+e^j}+E_{e^0+e^j}E_{e^0-e^j}\right),\end{array}\right.\nonumber
\end{eqnarray}
we have
\begin{eqnarray}\label{identitiesParta}
\left\{ \begin{array}{rcl}
\lambda_0^2+n\lambda_0 &= &c\cr
\lambda_1^2+(n-1)\lambda_1+\lambda_0 & = & c\cr
\lambda_2^2+(n-1)\lambda_2+(\lambda_1-\lambda_2)+\lambda_0 & = &
c\cr &\vdots &\cr
\lambda_n^2+(n-1)\lambda_n+(\lambda_1+\cdots+\lambda_{n-1}-(n-1)\lambda_n)+\lambda_0
& = & c.\end{array}\right.
\end{eqnarray}
Subtracting the 3rd identity from the 2nd identity, we have
$$
(\lambda_1-\lambda_2)(\lambda_1+\lambda_2+n-2)=0.
$$So $\lambda_1=\lambda_2=\lambda$ if
$n> 2$ and $\lambda_2=|\lambda_1|$ if $n=2$. Assume $n\ge 3$,
subtracting the 4th identity from the 3rd identity, we have
$$
(\lambda-\lambda_3)(\lambda+\lambda_2+n-3)=0.
$$ So $\lambda_1=\lambda_2=\lambda_3$ if
$n>3$ and $\lambda_1=\lambda_2=|\lambda_3|$ if $n=3$. By repeating
this argument $(n-1)$ times, we get
$\lambda_1=\cdots=\lambda_{n-1}=|\lambda_n|=\lambda$. Therefore, for
$n\ge 1$, we have
$|\Omega\rangle=|\lambda_0\underbrace{\lambda\cdots\lambda}_{n-1}(\pm\lambda)\rangle$.

By comparing the first two identities, we get
$$
(\lambda-\lambda_0)(\lambda+\lambda_0+n-1)=0.
$$ In view of the fact that $-\lambda_0\ge \lambda$, we conclude
that $(\lambda_0, \lambda)$ must be one of following pairs: 1)
(0,0), 2) $(-n-\lambda+1, \lambda )$ where $\lambda\ge 0$ is a half
integer. Consequently, when $n\ge 1$, the unitary highest weight
representation, if it exists, must be one of the following cases: 1)
the trivial one, 2) the one with highest weight $$(-(n-1+|\mu|),
\underbrace{|\mu|,\ldots, |\mu|}_{n-1},\mu)$$ for a half integer
$\mu$.

\vskip 10pt \noindent $\Leftarrow$: The remaining question we must
answer is this: such representations do exist and satisfy the
representation relations. This is certainly clear in the trivial
case.

The existence of such representations in the nontrivial case is
clear from the classification result of Refs. \cite{Jakobsen81a,
Jakobsen81b,EHW82}. As a matter of fact, in view of Theorem 1 in
Ref. \cite{MZ07a}, the one with highest weight $$(-(n-1+|\mu|),
\underbrace{|\mu|,\ldots, |\mu|}_{n-1},\mu)$$ can be realized by the
$(2n-1)$-dimensional generalized Kepler problem with magnetic charge
$\mu$. Moreover, in view of part 2) of Theorem 2 in Ref.
\cite{MZ07a}, these representations indeed satisfy the
representation relations. The only problem with this argument is
that the case $n=1$ is not covered; however, the results in Refs.
\cite{meng05, MZ07a} can be pushed down to the case $n=1$, see
subsection \ref{micz1} in appendix \ref{Appendix1}.

End of the proof of part 2) of Theorem \ref{RepTh2}.

\vskip 10pt We would like to remark that, by following the argument
in the proof of part 2) of Theorem \ref{RepTh1}, one can also verify
the representation relations directly. Since this argument is a bit
long, we choose to skip it.

\appendix

\section{MICZ-Kepler problems in dimensions one or
two}\label{Appendix1}

In Ref. \cite{meng05}, the generalized MICZ-Kepler problems are
introduced in dimension three or higher. Here we introduce their
limits in dimension one and dimension two. Since the arguments given
in Refs. \cite{meng05, MZ07a, M07} are still valid for these
limiting cases, the theorems listed below are stated without
detailed proof.

\subsection{MICZ-Kepler problems in dimension one}\label{micz1}

\begin{Def} Let $\mu$ a half integer and $|\mu|\ge 1/2$. Let ${\bb
R}_\mu$ be $\bb R_+$ if $\mu<0$ and be $\bb R_-$ if $\mu>0$. The
$1$-dimensional MICZ-Kepler problem with magnetic charge $\mu$ is
defined to be the quantum mechanical system on ${\bb R}_\mu$ for
which the wave-functions are complex-valued functions on ${\bb
R}_\mu$ and the hamiltonian is
\begin{eqnarray}
H
 = -\frac{1}{2}{d^2\over dx^2} + \frac{\mu^2-|\mu|}{2x^2}
 - \frac{1}{|x|}.
\end{eqnarray}
\end{Def}
Let $c=\mu^2-|\mu|$ and $p=-i{d\over d x}$. Define the dynamical
symmetry operators as follows:
\begin{eqnarray}
\left\{ \begin{array}{rcl} A &= &-{1\over 2}\left(xp^2+x+{c\over x}
\right),\cr M &= &-{1\over 2}\left(xp^2-x+{c\over x} \right)\cr T &=
& x p,\cr \Gamma &=& |x| p,\cr \Gamma_{-1} &= & {1\over
2}\left(|x|p^2+|x|+{c\over |x|} \right),\cr \Gamma_{2} &= & {1\over
2}\left(|x|p^2-|x|+{c\over |x|} \right).\end{array}\right.
\end{eqnarray}

Let the capital Latin letters $A$, $B$ run from $-1$ to $2$.
Introduce $J_{AB}$ as follows:
\begin{eqnarray}\label{defofJ1}
J_{AB}=\left\{\begin{array}{ll}  A & \hbox{if $A=1$, $B=2$}\cr M&
\hbox{if $A=1$, $B=-1$}\cr \Gamma & \hbox{if $A=1$, $B=0$}\cr T &
\hbox{if $A=2$, $B=-1$}\cr \Gamma_{2} & \hbox{if $A=2$, $B=0$}\cr
\Gamma_{-1} & \hbox{if $A=-1$, $B=0$}\cr -J_{BA} & \hbox{if
$A>B$}\cr 0 & \hbox{if $A=B$}.\cr
\end{array}\right.
\end{eqnarray}
The following theorem can be proved by direct computation:
\begin{Thm} Let $C^\infty({\bb R}_\mu)$ be the space of smooth complex-valued functions on ${\bb R}_\mu$. Let $J_{AB}$
be defined by (\ref{defofJ1}).

1) As operators on $C^\infty({\bb R}_\mu)$, $J_{AB}$'s satisfy the
following commutation relation:
\begin{eqnarray}[J_{AB},
J_{A'B'}]=-i\eta_{AA'}J_{BB'}-i\eta_{BB'}J_{AA'}+i\eta_{AB'}J_{BA'}+i\eta_{BA'}J_{AB'}\nonumber
\end{eqnarray}
where the indefinite metric tensor $\eta$ is ${\mr {diag}}\{++--\}$
relative to the following order: $-1$, $0$, $1$, $2$ for the
indices.

2) As operators on $C^\infty({\bb R}_\mu)$,
\begin{eqnarray} \{J_{AB},
{J^A}_C\}:=J_{AB}{J^A}_C+{J^A}_CJ_{AB}=2c\eta_{BC}.\nonumber
\end{eqnarray}
\end{Thm}

Consequently, one can obtain the following two theorems:
\begin{Thm}  For the $1$-dimensional
MICZ-Kepler problem with magnetic charge $\mu$, the following
statements are true:

1) The negative energy spectrum is
$$
E_I=-{1/2\over (I+|\mu|)^2}
$$ where $I=0$, $1$, $2$, \ldots;

2) The Hilbert space $\ms H$ of negative-energy states admits a
linear $\mr{Spin}(2)$-action under which there is a decomposition
$$
{\ms H}=\hat\bigoplus _{I=0}^\infty\,{\ms H}_I
$$ where ${\ms H}_I$ is the irreducible $\mr{Spin}(2)$-representation
witht weight $(I+|\mu|)\hbox{sign}(\mu)$;

3) $\mr{Spin}(1,1)$ acts linearly on the positive-energy states and
${\bb R}^{1}$ acts linearly on the zero-energy states;

4) ${\ms H}_I$ in part 2) is the energy eigenspace with eigenvalue
$E_I$ in part 1).
\end{Thm}

\begin{Thm}
Let ${\ms H}(\mu)$ be the Hilbert space of bound states for the
$1$-dimensional generalized MICZ-Kepler problem with magnetic charge
$\mu$.

1) There is a natural unitary action of $\mr{Spin}(2, 2)$ on $ {\ms
H}(\mu)$. In fact, ${\ms H}(\mu)$ is the unitary highest weight
module of $\mr{Spin}(2, 2)$ with highest weight $\left(-|\mu|,
\mu\right)$; consequently, it occurs at the unique reduction point
of the Enright-Howe-Wallach classification diagram\footnote{Page
101, Ref. \cite{EHW82}. See also Refs.
\cite{Jakobsen81a,Jakobsen81b}.} for the unitary highest weight
modules, so it is a non-discrete series representation.

2) As a representation of subgroup $\mr{Spin}(2, 1)$,
\begin{eqnarray}
{\ms H}(\mu) = {\mathcal D}^-_{2|\mu|}
\end{eqnarray} where ${\mathcal D}^-_{2|\mu|}$ is the anti-holomorphic discrete
series representation\footnote{The case $\mu=\pm 1/2$ is a limit of
the discrete series representation. } of $\mr{Spin}(2, 1)$ with
highest weight $-|\mu|$.

3) As a representation of the maximal compact subgroup
($=\mr{Spin}(2)\times_{\bb Z_2} \mr{Spin}(2)$),
\begin{eqnarray}
{\ms H}(\mu) = \hat \bigoplus_{l=0}^\infty D (-l-|\mu|)\otimes
D((l+|\mu|)\hbox{sign}(\mu))
\end{eqnarray} where $D(\lambda)$ denotes the irreducible module of $\mr{Spin}(2)$
with weight $\lambda$.

\end{Thm}

\subsection{MICZ-Kepler problems in dimension two}\label{micz2}
\begin{Def} Let $\mu = 0$ or $1/2$. The $2$-dimensional MICZ-Kepler problem with magnetic charge $\mu$
is defined to be the quantum mechanical system on $\bb R^{2}_*$ for
which the wave-functions are complex-valued functions $\psi $ on
$\bb R^{2}_*={\bb R}_+\times \bb R$ satisfying identity
$$\psi(r, \theta+2\pi)=(-1)^{2\mu}\psi(r, \theta)\quad \mbox{for any
$(r,\theta)\in {\bb R}_+\times \bb R$},$$ and the hamiltonian is
\begin{eqnarray}
H
 = -\frac{1}{2}\left({1\over r}\partial_rr\partial_r+{1\over r^2}{\partial^2\over \partial \theta^2}\right)
 - \frac{1}{r}.
\end{eqnarray}

\end{Def}
Let the small Greek letters $\alpha$, $\beta$ run from $1$ to $2$,
$x^1:=r\cos \theta$, $x^2:=r\sin \theta$, $x_\alpha:=x^\alpha$,
$p_\alpha:=-i{\partial\over
\partial x^\alpha}$. Define the dynamical symmetry operators as
follows:
\begin{eqnarray}\left\{
\begin{array}{rcl}
{ J}_{12}  &= & x_1 p_2 -x_2 p_1=-i\partial_\theta,\cr A_\alpha &=
&{1\over 2}x_\alpha p^2 - p_\alpha(\vec r\cdot \vec p) -{i\over
2}p_\alpha-{1\over 2}x_\alpha,\cr M_\alpha &= &{1\over 2}x_\alpha
p^2 - p_\alpha(\vec r\cdot \vec p) -{i\over 2}p_\alpha+{1\over
2}x_\alpha,\cr T &= & \vec r\cdot \vec p-{i\over 2},\cr
\Gamma_\alpha &=& r p_\alpha,\cr \Gamma_{-1} &= & {1\over
2}\left(rp^2+r \right),\cr \Gamma_{3} &= & {1\over 2}\left(rp^2-r
\right).\end{array}\right.
\end{eqnarray}

Let the capital Latin letters $A$, $B$ run from $-1$ to $3$.
Introduce $J_{AB}$ as follows:
\begin{eqnarray}\label{defofJ2}
J_{AB}=\left\{\begin{array}{ll}  J_{12} & \hbox{if $A=1$, $B=2$}\cr
A_\alpha & \hbox{if $A=\alpha$, $B=3$}\cr M_\alpha& \hbox{if
$A=\alpha$, $B=-1$}\cr \Gamma_\alpha & \hbox{if $A=\alpha$,
$B=0$}\cr T & \hbox{if $A=3$, $B=-1$}\cr \Gamma_{3} & \hbox{if
$A=3$, $B=0$}\cr \Gamma_{-1} & \hbox{if $A=-1$, $B=0$}\cr -J_{BA} &
\hbox{if $A>B$}\cr 0 & \hbox{if $A=B$}.\cr
\end{array}\right.
\end{eqnarray}

The following theorem can be proved by direct computation:
\begin{Thm} Let $C^\infty(\bb R^{2}_*)$ be the space of smooth complex-valued functions on $\bb R^{2}_*$. Let $J_{AB}$
be defined by (\ref{defofJ2}).

1) As operators on $C^\infty(\bb R^{2}_*)$, $J_{AB}$'s satisfy the
following commutation relation:
\begin{eqnarray}\label{cmtr} [J_{AB},
J_{A'B'}]=-i\eta_{AA'}J_{BB'}-i\eta_{BB'}J_{AA'}+i\eta_{AB'}J_{BA'}+i\eta_{BA'}J_{AB'}\nonumber
\end{eqnarray}
where the indefinite metric tensor $\eta$ is ${\mr {diag}}\{++---\}$
relative to the following order: $-1$, $0$, $1$, $2$, $3$ for the
indices.

2) As operators on $C^\infty(\bb R^{2}_*)$,
\begin{eqnarray} \{J_{AB},
{J^A}_C\}:=J_{AB}{J^A}_C+{J^A}_CJ_{AB}=-\eta_{BC}.\nonumber
\end{eqnarray}
\end{Thm}
Consequently, one can obtain the following two theorems:
\begin{Thm}  For the $2$-dimensional
MICZ-Kepler problem with magnetic charge $\mu$, the following
statements are true:

1) The negative energy spectrum is
$$
E_I=-{1/2\over (I+\mu+{1\over 2})^2}
$$ where $I=0$, $1$, $2$, \ldots;

2) The Hilbert space $\ms H$ of negative-energy states admits a
linear $\mr{Spin}(3)$-action under which there is a decomposition
$$
{\ms H}=\hat\bigoplus _{I=0}^\infty\,{\ms H}_I
$$ where ${\ms H}_I$ is the irreducible $\mr{Spin}(3)$-representation
with highest weight is $I+\mu$;

3) $\mr{Spin}(2,1)$ acts linearly on the positive-energy states and
$\mr{Spin}(2)\rtimes {\bb R}^{2}$ acts linearly on the zero-energy
states;

4) The linear action in part 2) extends the manifest linear action
of $\mr{Spin}(2)$, and ${\ms H}_I$ in part 2) is the energy
eigenspace with eigenvalue $E_I$ in part 1).

\end{Thm}
\begin{Thm}
Let ${\ms H}(\mu)$ be the Hilbert space of bound states for the
$2$-dimensional generalized MICZ-Kepler problem with magnetic charge
$\mu$.

1) There is a natural unitary action of $\widetilde{\mr{Spin}}(2,
3)$ on $ {\ms H}(\mu)$ which extends the manifest unitary action of
$\mr{Spin}(2)$. In fact, ${\ms H}(\mu)$ is the unitary highest
weight module of $\widetilde{\mr{Spin}}(2, 3)$ with highest weight
$\left(-(\mu+1/2),\mu\right)$; consequently, it occurs at the first
reduction point of the Enright-Howe-Wallach classification
diagram\footnote{Page 101, Ref. \cite{EHW82}. While there is a
unique reduction point when $\mu=1/2$, there are two reduction
points when $\mu=0$.  See also Refs.
\cite{Jakobsen81a,Jakobsen81b}.} for the unitary highest weight
modules, so it is a non-discrete series representation.

2) As a representation of $\mr{Spin}(2, 1)\times \mr{Spin}(2)$,
\begin{eqnarray}
{\ms H}(\mu) = \hat \bigoplus_{l=\mu+{\bb Z}}  {\mathcal
D}^-_{2|l|+1}\otimes D(l)
\end{eqnarray} where $D(l)$ is the irreducible module of $\mr{Spin}(2)$
with weight $l$ and ${\mathcal D}^-_{2|l|+1}$ is the
anti-holomorphic discrete series representation of $\mr{Spin}(2, 1)$
with highest weight $-(|l|+1/2)$.

3) As a representation of the maximal compact subgroup
($=\mr{Spin}(2)\times \mr{Spin}(3)$),
\begin{eqnarray}
{\ms H}(\mu) = \hat \bigoplus_{l=0}^\infty D(-(l+\mu+1/2))\otimes
D^l
\end{eqnarray} where $D^l$ is the irreducible module of $\mr{Spin}(3)$
with highest weight $l+\mu$ and $D(-(l+\mu+1/2))$ is the irreducible
module of $\mr{Spin}(2)$ with weight $-(l+\mu+1/2)$.

\end{Thm}


\begin{thebibliography}{99}

\bibitem{MC70}
H. McIntosh, A. Cisneros, Degeneracy in the presence of a magnetic
monopole, {\em J. Math. Phys.} {\bf 11} (1970), 896-916.

\bibitem{Z68}
D. Zwanziger, Exactly soluble nonrelativistic model of particles
with both electric and magnetic charges, {\em Phys. Rev.} {\bf 176}
(1968), 1480-1488.


\bibitem{Iwai90}
T. Iwai, The geometry of the $\mr{SU}(2)$ Kepler problem, {\em J.
Geom. Phys.} {\bf 7} (1990), 507-535.

\bibitem{meng05} G. W. Meng, MICZ-Kepler
problems in all dimensions. {\em J. Math. Phys.} {\bf 48} (2007),
032105. {\em E-print}, arXiv:math-ph/0507028.

\bibitem{Barut71}
A. Barut and G. Bornzin,  $\mr{SO}(4,2)$-Formulation of the Symmetry
Breaking in Relativistic Kepler Problems with or without Magnetic
Charges, {\em J. Math. Phys.} {\bf 12} (1971), 841-843.

\bibitem{MZ07a}
G. W. Meng and R. B. Zhang, Generalized MICZ-Kepler Problems and
Unitary Highest Weight Modules. {\em E-print},
arXiv:math-ph/0702086.

\bibitem{M07}
G. W. Meng, Generalized MICZ-Kepler Problems and Unitary Highest
Weight Modules -- II. {\em E-print}, arXiv:0704.2936.


\bibitem{EHW82} T. Enright, R. Howe and N. Wallach, A
classification of unitary highest weight modules, {\em
Representation theory of reductive groups}, Progress in Math. {\bf
40}, Birkh\"{a}user (1983), 97-143.


\bibitem{Jakobsen81a} H. P. Jakobsen, The last
possible place of unitarity for certain highest weight modules, {\em
Math. Ann.} {\bf 256} (1981), no. 4, 439-447.

\bibitem{Jakobsen81b}
H. P. Jakobsen, Hermitian symmetric spaces and their unitary highest
weight modules, {\em J. Funct. Anal.} {\bf 52} (1983), no. 3,
385-412.

\bibitem{BK96}
{\em Representation Theory and Automorphic Forms}, T.N. Bailey, A.W.
Knapp, eds, Proceedings of Symposia in Pure Mathematics, Vol. {\bf
61}, Amer. Math. Soc. 1997.


\bibitem{georgi82}
H. Georgi, {\em Lie Algebras in Particle Physics}, Benjamin, London
(1982).



\end{thebibliography}
\end{document}